\documentclass{PoS}

\usepackage[utf8]{inputenc}
\usepackage[T1]{fontenc}
\usepackage{newunicodechar}
\usepackage{graphicx} 
\usepackage{amsmath}
\usepackage{amsfonts,amsbsy}
\usepackage{amssymb}
\usepackage{subdepth}
\usepackage{bbold}
\usepackage{nicefrac}
\usepackage[dvipsnames]{xcolor}
\usepackage{mathrsfs}
\definecolor{lcolor}{rgb}{0.5,0,0}
\definecolor{citcolor}{rgb}{0,0.3,0.0}
\usepackage{braket, bbold, empheq, pdflscape, comment, pbox, leftidx}

\graphicspath{{Images/}}

\newcommand{\nc}{{N_\mathrm{c}}}

\newcommand{\as}{\alpha_\mathrm{s}}

\newcommand{\nr}[1]{(\ref{#1})}

\newcommand{\eq}{Eq.~}

\DeclareMathAlphabet{\ib}{OML}{cmm}{b}{it}

\newcommand{\uu}[2]{U_{\ib{{#1}}{{#2}}}}
\newcommand{\uub}[2]{\bar{U}_{\bar{\ib{{#1}}}{{#2}}}}
\newcommand{\ud}[2]{U^\dagger_{\ib{{#1}}{{#2}}}}

\newcommand{\ua}[3]{\tilde{U}^{{#1}}_{\ib{{#2}}{{#3}}}}
\newcommand{\uab}[3]{\bar{\tilde{U}}^{{#1}}_{\bar{\ib{{#2}}}{{#3}}}}
\newcommand{\tr}[1]{\mathrm{tr}\left\{{{#1}}\right\}}
\newcommand{\epr}[2]{e^{ i \epsilon g \alpha^R_{\ib{{{#1}}},{{#2}}}}}
\newcommand{\emr}[2]{e^{ - i \epsilon g \alpha^R_{\ib{{{#1}}},{{#2}}}}}
\newcommand{\epl}[2]{e^{ i \epsilon g \alpha^L_{\ib{{{#1}}},{{#2}}}}}

\newcommand{\ar}[2]{\alpha^R_{\ib{{{#1}}},{{#2}}}}
\newcommand{\al}[2]{\alpha^L_{\ib{{{#1}}},{{#2}}}}
\newcommand{\la}[3]{\lambda_{\ib{{#1}}{{#2}}}^{{#3}}}
\newcommand{\lab}[3]{\bar{\lambda}_{\bar{\ib{{#1}}}{{#2}}}^{{#3}}}

\newcommand{\ld}[2]{L^{{#1}}_{\ib{{{#2}}}}}
\newcommand{\rd}[2]{R^{{#1}}_{\ib{{{#2}}}}}
\newcommand{\ldb}[2]{\bar{L}^{{#1}}_{\bar{\ib{{{#2}}}}}}
\newcommand{\rdb}[2]{\bar{R}^{{#1}}_{\bar{\ib{{{#2}}}}}}
\newcommand{\ldalt}[3]{L^{{#1}}_{\ib{{{#2}}},#3}}
\newcommand{\rdalt}[3]{R^{{#1}}_{\ib{{{#2}}},#3}}
\newcommand{\ldbalt}[3]{\bar{L}^{{#1}}_{\bar{\ib{{{#2}}}},#3}}
\newcommand{\rdbalt}[3]{\bar{R}^{{#1}}_{\bar{\ib{{{#2}}}},#3}}

\title{Unequal rapidity correlators in the dilute limit of JIMWLK}

\ShortTitle{Unequal rapidity correlators}

\author{T. Lappi\\
        Department of Physics, %
 P.O. Box 35, 40014 University of Jyv\"askyl\"a, Finland\\
Helsinki Institute of Physics, P.O. Box 64, 00014 University of Helsinki,
Finland\\
        E-mail: \email{tuomas.v.v.lappi@jyu.fi}}

\author{\speaker{A. Ramnath}\\
        Department of Physics, P.O. Box 35, 40014 University of Jyv\"askyl\"a, Finland
\\
        E-mail: \email{andrecia.a.ramnath@student.jyu.fi}}

\abstract{We study unequal rapidity correlators in the stochastic Langevin picture of Jalilian-Marian--Iancu--McLerran--Weigert--Leonidov--Kovner (JIMWLK) evolution in the Color Glass Condensate effective field theory. By separately evolving the Wilson lines in the direct and complex conjugate amplitudes, we use the formalism to study two-particle production at large rapidity separations. We show that the evolution between the rapidities of the two produced particles can be expressed as a linear equation, even in the full nonlinear limit. We also show how the Langevin formalism for two-particle correlations reduces to a BFKL picture in the dilute limit and in momentum space, providing an interpretation of BFKL evolution as a stochastic process for color charges.}

\FullConference{XXVII International Workshop on Deep-Inelastic Scattering and Related Subjects - DIS2019\\
		8-12 April, 2019\\
		Torino, Italy}

\begin{document}

\section{Introduction}

The Color Glass Condensate (CGC, see e.g. \cite{Weigert:2005us,Gelis:2010nm}) is an effective theory of QCD for high energy processes. The JIMWLK\footnote{
	The acronym stands for Jalilian-Marian--Iancu--McLerran--Weigert--Leonidov--Kovner.} evolution equation~\cite{Jalilian-Marian:1997xn,Jalilian-Marian:1997gr,Iancu:2001md,Ferreiro:2001qy,Mueller:2001uk}, 
can be used to resum leading logarithmic (in energy or $x$) corrections to QCD scattering cross sections. In addition to providing a more direct physical picture of the evolution, the Langevin formulation is the basis for numerical solutions of the JIMWLK equation~\cite{Kovchegov:2008mk,Lappi:2012vw}.

The most common phenomenological applications of the CGC framework involve processes in which one needs only the Wilson lines at one rapidity. The situation becomes more complicated if one is interested in the correlations between particles that are separated by a parametrically large rapidity interval $\Delta Y \gtrsim 1/\as$. For this purpose, a formalism based on the Langevin description of JIMWLK evolution was developed by Iancu and Triantafyllopoulos (IT) in \cite{Iancu:2013uva} (see also earlier, very similar work in \cite{Kovner:2006ge,Kovner:2006wr}). Our intention in this paper, following the more detailed discussion in \cite{Lappi:2019kif},  is to analyze this further.

\section{JIMWLK evolution and particle production at equal rapidity}
\label{sec:jimwlk}

We consider a high energy interaction of a dilute colored probe with the color field of a dense target. 
The expectation value of an observable $\hat{\mathcal{O}}$ is given by $\left\langle \hat{\mathcal{O}} \right\rangle_Y \equiv \int [DU] W_Y[U] \hat{\mathcal{O}}$, where $W_Y[U]$ is the CGC weight function describing the density distribution at $Y$ of the Wilson lines $\ud{x}{} \equiv P \exp \left\{ i g \int dx^+ \alpha^a_\ib{x}(x^+) t^a \right\}$ in the target.
The dependence of the target color field on rapidity is described by JIMWLK evolution.  The CGC weight function evolves from an initial condition $Y_\mathrm{in}$ to a final $Y$ according to the JIMWLK equation $\frac{\partial}{\partial Y} W_Y [U] = H W_Y[U]$.
The JIMWLK Hamiltonian is
$H \equiv \frac{1}{8 \pi^3} \int _\ib{uvz} \mathcal{K}_\ib{uvz} (\ld{a}{u} - \ua{\dagger ab}{z}{} \rd{b}{u}) (\ld{a}{v} - \ua{\dagger ac}{z}{} \rd{c}{v})$,
where tildes denote the adjoint representation.
The JIMWLK kernel is $\mathcal{K}_\ib{uvz} \equiv \mathcal{K}^i_\ib{uz} \mathcal{K}^i_\ib{vz}$,
where $\mathcal{K}^i_\ib{uz} = \frac{(\ib{u} - \ib{z})^i}{(\ib{u} - \ib{z})^2}$
is the Weizs\"acker-Williams soft gluon emission kernel. 
The $L$ and $R$ are ``left'' and ``right'' Lie derivatives that act to color-rotate the Wilson lines. 
They are defined as
$\ld{a}{u} \equiv - i g (\uu{u}{} t^a)_{\alpha\beta} \frac{\delta}{\delta \uu{u}{,\alpha\beta}}$ and
$\rd{a}{u} \equiv - i g (t^a \uu{u}{})_{\alpha\beta} \frac{\delta}{\delta \uu{u}{,\alpha\beta}}$.

In the Langevin formulation, evolution is treated as a random walk in the functional space of Wilson lines. 
Rapidity is discretized as $Y - Y_0 = \epsilon N$ with $\mathbb{Z} \ni N \rightarrow \infty, \epsilon \rightarrow 0$, where each evolution step is labelled by $n \in \{0,1,...,N\}$. 
The noise is introduced within terms we can call, respectively, ``left'' and ``right'' (traceless, Hermitian) color fields
$\al{x}{n}
\equiv \int_\ib{z} \mathcal{K}^i_\ib{xz} \nu^i_{\ib{z},n}/\sqrt{4\pi^3}$
and
$\ar{x}{n} 
\equiv \int_\ib{z} \mathcal{K}^i_\ib{xz} \uu{z}{,n} \nu^i_{\ib{z},n} \ud{z}{,n}/\sqrt{4\pi^3}$,
where $\nu^i_{\ib{z},n} \equiv \nu^{i,a}_{\ib{z},n} t^a$. $\nu^{i,a}_{\ib{z},m} \in \mathbb{R}$ and satisfies $\left\langle \nu^{i,a}_{\ib{x},m} \nu^{j,b}_{\ib{y},n} \right\rangle = \frac{1}{\epsilon} \delta^{ij} \delta^{ab} \delta_{mn} \delta_\ib{xy}$. The Langevin equation describing the evolution of a Wilson line is
$\ud{x}{,n+1} = \epl{x}{n} \ud{x}{,n} \emr{x}{n}$.
Since $\epsilon$ is infinitesimal, we may use it as an expansion parameter to obtain
\begin{align}
 \ud{x}{,n+1}
=& \; \ud{x}{,n} 
+ \int_\ib{z} \left( \frac{i \epsilon g}{\sqrt{4\pi^3}} \mathcal{K}^i_\ib{xz} \nu^{i,a}_{\ib{z},n} - \frac{\epsilon g^2}{4\pi^3} \mathcal{K}_\ib{xxz} t^a \right)
(t^a \ud{x}{,n} - \ud{x}{,n} \ua{\dagger ab}{z}{,n} t^b)
+ \mathcal{O}(\epsilon^{3/2}).
\label{ud-eps}
\end{align}
The Balitsky-Kovchegov~\cite{Balitsky:1995ub, Kovchegov:1999yj} (BK) equation can be obtained from this by first calculating the dipole $\hat{S}_{\ib{x\bar{x}},n+1}$, then using the Fierz identity and finally taking the mean field approximation.

Next, consider a single quark produced in a proton-nucleus collision. It is described mathematically by a fundamental representation dipole $\hat{S}_\ib{x\bar{x}} \equiv \tr{U^\dagger_\ib{x} \bar{U}_\ib{\bar{x}}}/\nc$. The bars on both the Wilson line and the coordinate in $\bar{U}_\ib{\bar{x}}$ denote that this Wilson line is in the CCA. 
The cross section for inclusive quark production in a proton-nucleus collision is then 
\begin{align}
\frac{d\sigma_{q}}{d\eta_p d^2\ib{p}} = x q (x) \frac{1}{(2 \pi)^2} \int_\ib{x\bar{x}} e^{-i \ib{p} \cdot (\ib{x} - \ib{\bar{x}})} \left\langle \left. \hat{S}_\ib{x\bar{x}} \right|_{\bar{U} = U} \right\rangle_Y.
\end{align}
Here, $Y$ is the relative rapidity of the produced quark with respect to the target, $x$ is the longitudinal momentum fraction of the projectile, $x q(x)$ is the quark distribution in the proton, and $\ib{p}$ and $\eta_p$ are the transverse momentum and rapidity, respectively, of the quark. 

For inclusive quark-gluon production, the cross section can be written compactly in terms of a ``production Hamiltonian'' \cite{Kovner:2006ge,Kovner:2006wr,Iancu:2013uva} operating on the quark cross section:
\begin{align}
\frac{d\sigma_{qg}}{d\eta_p d^2\ib{p} \, d\eta_k d^2\ib{k}} =
\frac{1}{(2 \pi)^4} \int_\ib{x\bar{x}} e^{-i \ib{p} \cdot (\ib{x} - \ib{\bar{x}})} \left\langle \left. H_\mathrm{prod}(\ib{k}) \hat{S}_\ib{x\bar{x}} \right|_{\bar{U} = U} \right\rangle_Y.
\label{cross-sec}
\end{align}
Here, the quark has transverse momentum $\ib{p}$ and pseudo-rapidity $\eta_p$, and the gluon has transverse momentum $\ib{k}$ and pseudo-rapidity $\eta_k$.
The production Hamiltonian is given by~\cite{Iancu:2013uva}
$H_\mathrm{prod} (\ib{k}) = \frac{1}{4\pi^3} \int _\ib{y \bar{y}} e^{-i \ib{k} \cdot (\ib{y} - \ib{\bar{y}})} \int_\ib{u\bar{u}} \mathcal{K}^i_\ib{yu} \mathcal{K}^i_\ib{\bar{y} \bar{u}}
(\ld{a}{u} - \ua{\dagger ab}{y}{} \rd{b}{u}) (\ldb{a}{u} - \uab{\dagger ac}{y}{} \rdb{c}{u})$.

\section{Dilute limit: stochastic picture of BFKL evolution}
\label{sec:dilute}

We start with the fundamental representation Wilson line
$\ud{x}{,n} 
= e^{i \la{x}{,n}{}}
= \mathbb{1} + i \la{x}{,n}{} - \frac{1}{2} \la{x}{,n}{2} + \mathcal{O}(\la{}{}{3})$,
where each real matrix $\lambda$ is an element of the algebra of SU$(\nc)$ and denotes a one-gluon interaction between projectile and target. The full Langevin step to linear order is then
\begin{multline}
\la{x}{,n+1}{}
= \la{x}{,n}{}
+ \int_\ib{z} \left( \frac{i \epsilon g}{\sqrt{4\pi^3}} \mathcal{K}^i_\ib{xz} \nu^{i,a}_{\ib{z},n} - \frac{\epsilon g^2}{4\pi^3} \mathcal{K}_\ib{xxz} t^a \right) 
i f^{abc} t^c (\la{x}{,n}{b} - \la{z}{,n}{b})
+ \mathcal{O}(\epsilon^{3/2}, \la{}{}{2}).
\label{langevin-dilute}
\end{multline}

The BFKL equation can be obtained by first expanding the Wilson lines in the dilute limit, and looking at the evolution of a quantity that is quadratic in the expansion parameter $\lambda$. We first square \eq\nr{langevin-dilute} to obtain an iterative equation for $\la{x}{}{a} \lab{x}{}{a}$. From this basic equation, one can define two different versions of the BFKL equation. For the first, we define the unintegrated gluon distribution $\phi^n_{\ib{x\bar{x}}} \equiv \langle \la{x}{,n}{a} \lab{x}{,n}{a} \rangle$.
After Fourier transforming, $\la{x}{,n+1}{a} \lab{x}{,n+1}{a}$ takes the familiar form of
\begin{align}
\phi^{n+1}({\ib{q}})
= \phi^n({\ib{q}})
+ 4 \nc \epsilon \alpha_s \int_\ib{p} \frac{1}{\ib{(q - p)}^2} \bigg(
\frac{\phi^n({\ib{p}}) \ib{p}^2}{\ib{q}^2}
- \frac{1}{2} \frac{\phi^n({\ib{q}}) \ib{q}^2}{\ib{p}^2} \bigg)
+ \mathcal{O}(\epsilon^{3/2}, \phi^{3/2}), \nonumber
\end{align}
the (color singlet, zero momentum transfer) textbook version of the BFKL equation \cite{Forshaw:1997dc}. 

The other (Mueller's) version of the BFKL equation~\cite{Mueller:1994jq} is obtained when one looks at the expansion of the dipole operator as $\tr{\ud{x}{} \uu{y}{}}/\nc = 1 - \frac{1}{4 \nc} (\la{x}{}{a} - \la{y}{}{a}) (\la{x}{}{a} - \la{y}{}{a}) + \mathcal{O}(\la{}{}{3})$.
The natural definition of the gluon distribution based on this expansion is then the so-called ``BFKL pomeron'' \cite{Caron-Huot:2013fea} $\varphi_\ib{xy} \equiv \left< (\la{x}{}{a} - \la{y}{}{a}) (\la{x}{}{a} - \la{y}{}{a})\right>$,
which we can write in terms of $\phi$ by setting $\bar{\lambda} = \la{}{}{}$: $\phi_\ib{xx}+ \phi_\ib{yy} - 2\phi_\ib{xy} \overset{\bar{\lambda} = \la{}{}{}}{=} \varphi_\ib{xy}$.
One then arrives at the Mueller version of the BFKL equation:
$\varphi^{n+1}_{\ib{xy}} - \varphi^n_{\ib{xy}} =
- \frac{\nc}{2} \frac{\epsilon \alpha_s}{\pi^2} \int_\ib{z} 
 \tilde{\mathcal{K}}_\ib{xyz}
[\varphi^n_{\ib{xy}} - \varphi^n_{\ib{xz}} - \varphi^n_{\ib{zy}}]$.

\section{Unequal rapidity correlators in JIMWLK}
\label{sec:neqy}

Next, we want to calculate the double inclusive cross section for the simultaneous production two particles, separated in rapidity such that $\as(Y-Y_A) \gg 1$. The expectation value of the cross section for producing a quark at some rapidity $Y$ is calculated as an average over the noise $\nu$ at the end of the stochastic process: $\left\langle \hat{S}_{\ib{x\bar{x}}} \right\rangle_{Y-Y_A} = \left\langle \hat{S}_{\ib{x\bar{x}},N} \right\rangle_\nu$.
For the expectation value of an operator at the later rapidity $Y$, we now have
$\left\langle \hat{\mathcal{O}} 
\right\rangle_{Y - Y_A} 
\equiv \int [DU D\bar{U}] W_{Y - Y_A} [U, \bar{U}| \uu{}{A}, \bar{U}_A] \hat{\mathcal{O}}$.
We need to have a new conditional weight function $W_{Y - Y_A} [U, \bar{U}| \uu{}{A}, \bar{U}_A]$ \cite{Gelis:2008sz}, which obeys the differential equation 
$\frac{\partial}{\partial Y} W_{Y - Y_A} [U, \bar{U}| \uu{}{A}, \bar{U}_A] = H_\mathrm{evol} W_{Y - Y_A} [U, \bar{U}| \uu{}{A}, \bar{U}_A]$. The initial condition at $Y_A$ for the conditional weight function sets Wilson lines for both the DA and the CCA:
$W_{Y_A} [U, \bar{U}| \uu{}{A}, \bar{U}_A] = \delta [U - \uu{}{A}] \delta [\bar{U} - \bar{U}_A]$.

For a gluon emitted from a quark projectile at rapidity $Y_A$, we must now operate with the production Hamiltonian \emph{acting on the Wilson lines at} $Y_A$ \cite{Iancu:2013uva}. The final result is
\begin{multline}
\frac{d\sigma_{qg}}{dY d^2\ib{p} \, dY_A d^2\ib{k}_A}
= \frac{1}{(2 \pi)^4} \frac{1}{4\pi^3} \frac{1}{\nc} \int_\ib{x\bar{x}y\bar{y}} e^{-i \ib{p} \cdot (\ib{x} - \ib{\bar{x}})} 
e^{-i\ib{k}_A \cdot (\ib{y} - \ib{\bar{y}})} 
\int_\ib{uv} 
\mathcal{K}^i_\ib{yu} \mathcal{K}^i_\ib{\bar{y}v} \left\langle \left\langle \mathcal{I}_N \right\rangle_\ib{\nu} \right\rangle_{Y_A},
\label{final} 
\end{multline}
\begin{multline}
\mathcal{I}_n := \tr{\ldbalt{a}{u}{0} \uub{x}{,n} \ldalt{a}{u}{0} \ud{x}{,n}}
- \uab{\dagger ac}{y}{,0} \tr{\rdbalt{c}{u}{0} \uub{x}{,n} \ldalt{a}{u}{0} \ud{x}{,n}}
- \ua{\dagger ab}{y}{,0} \tr{\ldbalt{a}{u}{0} \uub{x}{,n} \rdalt{b}{u}{0} \ud{x}{,n}}
\\
+ \ua{\dagger ab}{y}{,0} \uab{\dagger ac}{y}{,0} \tr{\rdbalt{c}{u}{0} \uub{x}{,n} \rdalt{b}{u}{0} \ud{x}{,n}}.
\label{mathcal-i}
\end{multline}
To find the expressions for $R\ud{}{}, R\uu{}{}, L\ud{}{}$ and $L\uu{}{}$, one acts with the Lie derivatives on the evolution equations for the Wilson lines. However, the four equations are not independent of each other. For example, we may start by finding the equation for $R\ud{}{}$. The Hermitian conjugate will give the equation for $R\uu{}{}$, and the relation $\ldalt{a}{u}{0} = \ua{\dagger ab}{u}{,0} \rdalt{b}{u}{0}$ can be used to get the equations for $L\ud{}{}$ and $L\uu{}{}$.
Instead of an equation for $\rdalt{a}{u}{0} \ud{x}{,n+1}$, it is  more natural to define a quantity $R^a_{\ib{ux},n} \equiv \uu{x}{,n} \rdalt{a}{u}{0} \ud{x}{,n}$, so we can write the Langevin step compactly as
\begin{align}
R^a_{\ib{ux},n+1} = \epr{x}{n} R^a_{\ib{ux},n} \emr{x}{n}
- \frac{i \epsilon g}{\sqrt{4\pi^3}} \epr{x}{n} \int_\ib{z} \mathcal{K}^i_\ib{xz} [\tilde{\nu}^i_{\ib{z},n}, R^a_{\ib{uz},n}],
\label{ru-dagger2}
\end{align}
where $\tilde{\nu}^i_{\ib{z},n} = \uu{z}{,n} \nu^i_{\ib{z},n} \ud{z}{,n}$. This is linear and independent of the Wilson lines, and we can therefore express the evolution between the two rapidities in terms of linear BFKL-like dynamics. The whole cross section, however, is not given by a ``$k_T$-factorized'' expression (unlike the dilute case that we discuss in the next section), due to the explicit appearance of the Wilson lines in the cross section. 

\section{Two-particle correlators in the dilute limit}
 \label{sec:dilute2part}

The essential part of the cross section \nr{final} is given by $\mathcal{I}_n$ as defined in \eq\nr{mathcal-i}. This requires the operations of the Lie derivatives in the dilute limit. Using this, we obtain an equation for $\rdalt{a}{u}{0} \la{x}{,n+1}{}$. The equation for $L\ud{}{}$ is identical, with $R \rightarrow L$.
Combining these into the linearized production Hamiltonian and the linearized dipole, we get
\begin{align}
\mathcal{I}_n
= \frac{g^2}{2 \nc} f^{abc} f^{ade} (\lab{u}{,0}{e} - \lab{y}{,0}{e}) (\la{u}{,0}{c} - \la{y}{,0}{c}) 
\frac{\delta}{\delta \lab{u}{,0}{d}} \frac{\delta}{\delta \la{u}{,0}{b}}
\lab{x}{,n}{f} \la{x}{,n}{f} + \mathcal{O}(\la{}{}{3}).
\end{align}
 
The relation between $\la{x}{,n}{b}$ and $\la{u}{,0}{a}$ is linear. Thus the Green's function, defined by $\mathcal{F}^{n}_{\ib{x,\bar{x},u,\bar{u}}} \equiv 
\frac{\delta}{\delta \lab{u}{,0}{a}} \frac{\delta}{\delta \la{u}{,0}{a}}
\lab{x}{,n}{b} \la{x}{,n}{b}$, does not depend on $\lambda$. Using this we get 
$\langle \mathcal{I}_N \rangle
= \frac{g^2}{2} (\phi^0_\ib{\bar{u} u} - \phi^0_\ib{\bar{u} y} - \phi^0_\ib{\bar{y} u} + \phi^0_\ib{\bar{y} y})
\mathcal{F}^N_\ib{x,\bar{x},u,\bar{u}} + \mathcal{O}(\phi^{3/2})$ in terms of the gluon distribution, 
which we can put into the equation for the two-particle cross section to obtain a $k_T$-factorized expression:
\begin{align}
\frac{d\sigma_{qg}}{dY d^2\ib{p} \, dY_A d^2\ib{k}_A} = 
\frac{1}{(2 \pi)^4} \frac{1}{2 \nc} \frac{\alpha_s}{\pi^2} \int_\ib{x\bar{x}y\bar{y}u\bar{u}} 
\mathcal{K}^i_\ib{yu} \mathcal{K}^i_\ib{\bar{y}\bar{u}}
e^{-i \ib{p} \cdot (\ib{x} - \ib{\bar{x}}) -i\ib{k}_A \cdot (\ib{y} - \ib{\bar{y}})} \nonumber
\\
\times (\phi^0_\ib{\bar{u} u} - \phi^0_\ib{\bar{u} y} - \phi^0_\ib{\bar{y} u} + \phi^0_\ib{\bar{y} y})
\mathcal{F}^N_\ib{x,\bar{x},u,\bar{u}} + \mathcal{O}(\phi^{3/2}).
\label{eq:cspfact}
\end{align}
We can finally take the color field to be equal in the DA and the CCA. The fact that we are taking derivatives with respect to $\lab{u}{,0}{}$ and $\la{u}{,0}{}$ does not interfere with the BFKL evolution of the gluon density $\lab{x}{,n}{b} \la{x}{,n}{b}$. So $\mathcal{F}^{n}_{\ib{x,\bar{x},u,\bar{u}}}$ satisfies the same equation as $\lab{x}{,n}{b} \la{x}{,n}{b}$ with respect to index $n$.

Fourier transforming everything, we get
\begin{align}
\frac{d\sigma_{qg}}{dY d^2\ib{p} \, dY_A d^2\ib{k}_A} =
- \frac{\alpha_s}{\nc} \int_\ib{q} \frac{\ib{q}^2}{(\ib{q} - \ib{k}_A)^2 \ib{k}_A^2}  \mathcal{F}^{N}(\ib{-p,p,\ib{q}-k_A,-q+k_A}) \varphi^0(-\ib{q})
+ \mathcal{O}(\varphi^{3/2}). \nonumber
\end{align}
with the (zero momentum transfer) BFKL Green's function $\mathcal{F}$ satisfying the usual BFKL equation. We have thus shown that the IT Langevin equation formalism reduces, in the dilute limit, to a conventional correlation between two particles produced from the same  BFKL ladder. The initial condition for the evolution is $\mathcal{F}^0(\ib{P,\bar{P},m,\bar{m}})
= (\nc^2 -1) \delta^{(2)}(\ib{P}+\ib{m}) \delta^{(2)}(\ib{\bar{P}}+\ib{\bar{m}})$.
Using this in the general expression reduces the equal rapidity cross section to a $k_T$-factorized expression:
\begin{equation}
\left. 
\frac{d\sigma_{qg}}{dY d^2\ib{p} \, dY_A d^2\ib{k}_A}
\right|_{Y=Y_A}
= 
- \frac{\as}{(2 \pi)^2} \frac{(\ib{p}+\ib{k}_A)^2}{\ib{p}^2\ib{k}_A^2}
\varphi^0(\ib{p}+\ib{k}_A).
\label{eq:equalrap}
\end{equation}

\section{Conclusions}

We have attempted to clarify the Langevin formulation~\cite{Iancu:2013uva} of two-particle correlations in JIMWLK evolution, in the case of a dilute probe scattering off a dense color field target. Although JIMWLK evolution for the Wilson lines is nonlinear, the evolution of the Lie derivatives encoding the correlation between the two rapidities, is in fact not. It can be expressed as a linear equation that is independent of the Wilson lines. This observation seems to confirm the result obtained earlier (in a rather different language) in \cite{JalilianMarian:2004da}. 

We have also calculated explicitly the dilute limit of the Langevin formulation, where the decorrelations in azimuthal angle between the two particles are given by a BFKL Green's function between the two rapidities. JIMWLK evolution as a function of the quark rapidity $Y$ ``commutes'' with the production Hamiltonian and only operates on the dipole operator at $Y$. The evolution of the double inclusive cross section with $Y$ is therefore determined by the evolution of the single inclusive cross section, but with a more complicated initial condition.

\begin{acknowledgments}
We are grateful to R. Boussarie, M. Lublinsky, E. Iancu and D. Triantafyllopoulos for discussions.
 T.~L. has been supported by the Academy of Finland, projects No. 267321 and No. 303756. A.~R. is supported by the National Research Foundation of South Africa. This work has been supported by the European Research Council, grant ERC-2015-CoG-681707.
\end{acknowledgments}

\bibliography{spires}

\end{document}